\documentclass[twocolumn,showpacs,amsmath,amssymb,prl,graphics,graphicx
]{revtex4-1}
\usepackage{bm}
\usepackage[pdftex]{graphicx}
\usepackage{dcolumn}
\begin{document}

\title{Effect of equatorial  line nodes on upper critical field and London penetration depth}
\author{V. G. Kogan  and R. Prozorov}
\affiliation{Ames Laboratory - DOE and Department of Physics,
             Iowa State University, Ames, IA 50011}
             
            \begin{abstract}
            The upper critical field $H_{c2}$ and its anisotropy are calculated for  order parameters with line nodes at  equators, $k_z=0$, of the Fermi surface of uniaxial superconductors. It is 
  shown that characteristic features found in Fe-based materials -- a nearly linear $H_{c2}(T)$   in a broad $T$ domain, a low and increasing on warming anisotropy  $\gamma_H= H_{c2,ab}/ H_{c2,c}\, $ -- can be caused by competing effects of the equatorial nodes and of the Fermi surface anisotropy. For certain material parameters, $\gamma_H(T)-1$  may change sign on warming
in agreement with   recorded behavior of FeTeS system. It is also shown that the anisotropy of the penetration depth $\gamma_\lambda= \lambda_c/\lambda_{ab} $ decreases on warming to reach $\gamma_H$ at $T_c$ in agreement with data available. For some materials $\gamma_\lambda(T)$ may change on warming from $\gamma_\lambda>1$ at  low $T$s to  $\gamma_\lambda<1$ at  high $T$s.  
\end{abstract}



\date{9 May 2014}  

\maketitle 

 Iron-based superconductors are layered compounds with nearly two-dimensional Fermi surfaces
which at first sight  should have lead to high anisotropies of the upper critical field and the London penetration depth. This, however, is not the case. Most of these compounds have relatively low values of $\gamma_H=H_{c2,ab}/H_{c2,c}$ that {\it increase on warming} \cite{Altar} and in some materials even change from $\gamma_H<1$ at low temperatures to $\gamma_H>1$ at high $T$s \cite{FeTeS1,FeTeS}. The anisotropy of the London penetration depth is also low but {\it decreases on warming} \cite{gam-lam(t)}. Originally, such behavior was attributed to  multiband physics similar to the  two-band MgB$_2$ \cite{MgB2}. However, in MgB$_2$,   $\gamma_H (T)$ decreases on warming whereas $\gamma_\lambda(T)$ increases, i.e., just the opposite to Fe-based materials. 

Recently, the increasing $\gamma_H(T)$ had been associated with 
 the order parameter modulated along the c-axis \cite{KP-ROPP} even in the single-band scenario, so that multi-band effects {\it per se} are not necessary to  explain  the observations. It is also known that some iron-based superconductors have gap nodes and there are models suggesting equatorial line nodes \cite{theory-nodes}. Such gap structure is seen in the ARPES data  on BaFe$_2$(As$_{0.7}$P$_{0.3}$)$_2$ \cite{Feng} and was also explored for other unconventional superconductors, for example, Sr$_2$RuO$_4$ to understand anisotropic thermal conductivity  \cite{Tanatar,Mackenzie}.

We show in this Letter that the competing effects of equatorial nodes and the Fermi surface anisotropy might be responsible for the observed behavior of $H_{c2}$ in these materials. Moreover, we show that equatorial line nodes may cause the anisotropy of the London penetration depth, $\gamma_\lambda$, to {\it decrease on warming},  the feature seen in a number of materials for which data on $\lambda$-anisotropy are available \cite{gam-lam(t)}. The interplay of the Fermi surface effects and those due to line nodes can result in the temperature dependent sign of $\gamma_\lambda-1$, the prediction to be verified. In particular we show that this interplay may cause the in-plane superfluid density   to change with temperature in a ``d-wave-like" fashion (linear at low $T$s) while being rather flat at low $T$s for the $c$ direction reminiscent of the ``s-wave" manner.  \\
 
Studying the orbital $H_{c2}(T)$, we employ a version of Helfand-Werthamer (HW) theory \cite{HW}  generalized for clean anisotropic superconductors   \cite{KP-ROPP}. It is based on Eilenberger quasi-classical formulation of the  superconductivity \cite{Eil} with a weak-coupling separable potential   $V(\bm k,\bm k^\prime)=V_0\Omega(\bm k)\Omega(\bm k^\prime)$ and the order parameter in the form   $\Delta = \Psi({\bf r},T)\,\Omega({\bf k})$, $\bm k$ is the Fermi momentum \cite{Kad}.   $ \Omega({\bf k})$  determines the $\bm k$ dependence of $\Delta$ and is normalized so that the average over the Fermi surface 
$\langle\Omega^2\rangle=1$.   This popular approximation  works well for one band materials with anisotropic coupling and can be generalized to a multi-band case \cite{KP-ROPP}.  

Within this theory, $H_{c2,c}$ along the $c$ axis of uniaxial crystals is found by solving an equation \cite{KP-ROPP}:
 \begin{eqnarray}
&&   \ln t= 2 h_c \int_0^{\infty}s\, \ln\tanh (st) \,
 \left \langle\Omega^2 \mu_c  e^{-\mu_c h_c s^2 }\right\rangle ds\,, \qquad
  \label{eq-hc}\\
&& h_c=H_{c2,c}\frac{\hbar^2v_0^2}{2\pi \phi_0T_c^2},\,\, \mu_c=\frac{v_x^2+v_y^2}{v_0^2} ,\,\, v_{0 }^3= \frac{2E_F^2}{\pi^2\hbar^3N(0) }  .\qquad
 \label{mu_c}
\end{eqnarray}
Here, $t=T/T_c$, 
$v_x,v_y$ are Fermi velocities in the $a,b$ plane,  $E_F$ is the Fermi energy, and $N(0)$ is the total density of states at the Fermi level per spin. One easily verifies that the velocity $v_0=v_F$ for the isotropic case.

In principle, Eq.\,(\ref{eq-hc}) can  be used to evaluate $h_c(t)$ for any order parameter anisotropy (any $\Omega$) and any Fermi surface (any $\mu_c$).
 Both $\Omega$ and $\mu_c$ enter Eq.\,(\ref{eq-hc}) under the sign of the Fermi surface averaging and one does not expect fine details of  Fermi surface to affect strongly the  $H_{c2,c}(T)$ shape. In fact, this is what made the isotropic HW model  so successful. For this reason, describing   Fermi surface shapes,  we focus on a simplest version of Fermi spheroids, for which the averaging is a well defined analytic procedure, see e.g. \cite{MKM,KP-ROPP}.  

In general, Eq.\,(\ref{eq-hc}) can be solved numerically, but if $T\to T_c$,  the result is exact \cite{KP-ROPP}:
\begin{eqnarray}
  h_c= \frac{8(1-t)}{7\zeta(3)  \left\langle \Omega^2\mu_c \right\rangle } \,,\quad   h_c^\prime(1)= -\frac{8 }{7\zeta(3)  \left\langle \Omega^2\mu_c \right\rangle } \,.\qquad
    \label{hc(Tc)}
 \end{eqnarray}
Here, $\zeta(3)\approx1.202$, and $h_c^\prime(1)=(dh_c/dt)_{t=1}$. 
For the isotropic case with $\Omega=1$ and $\mu_c=2/3$, one reproduces the  HW slope  near $T_c$ in the clean limit. 

At $T=0$, Eq.\,(\ref{eq-hc}) was shown to yield  \cite{KP-ROPP}: 
 \begin{eqnarray}
 h_c(0)=e^{ -\bm C -  \langle\Omega^2 \ln \mu_c      \rangle } \,,
  \label{h_c(0)}
\end{eqnarray}
where $\bm C\approx 0.577$ is the Euler constant.  Hence, we obtain the HW ratio, 
 \begin{eqnarray}
   h_c^*(0) =\frac{H_{c2,c}(0)}{T_c H_{c2,c}^\prime (T_c)}= \frac{ h_c(0)}{h_c^\prime(1)} = \frac{7\zeta(3)}{8e^{\bm C}}\, \langle\Omega^2\mu_c\rangle e^{-\langle\Omega^2\ln\mu_c\rangle };\qquad
\label{HW}
\end{eqnarray}
 $H_{c2,c}^\prime(T_c)\equiv dH_{c2,c}/dT$ at $T_c$. 
For the isotropic case this gives the clean limit HW value 
  $ h_c^*(0) =   7\zeta(3)/48e^{\bm C-2}=0.727$.

Thus,  both the order parameter symmetry and the Fermi surface affect $h_c^*(0)$.
It is worth noting, however, that for s-wave order parameters on Fermi spheroids, $h_c^*(0)$ remains close to   $0.7$ independently of  the ratio of the  spheroid semi-axes  \cite{KP-ROPP}. We also note that   $h_c^*(0)$ is nearly insensitive to the non-magnetic transport scattering, but it  decreases fast in the presence of pair breaking to reach 0.5 for the strong $T_c$ suppression \cite{KP}. \\
 \begin{figure}[b]
\begin{center}
 \includegraphics[width=8.5cm] {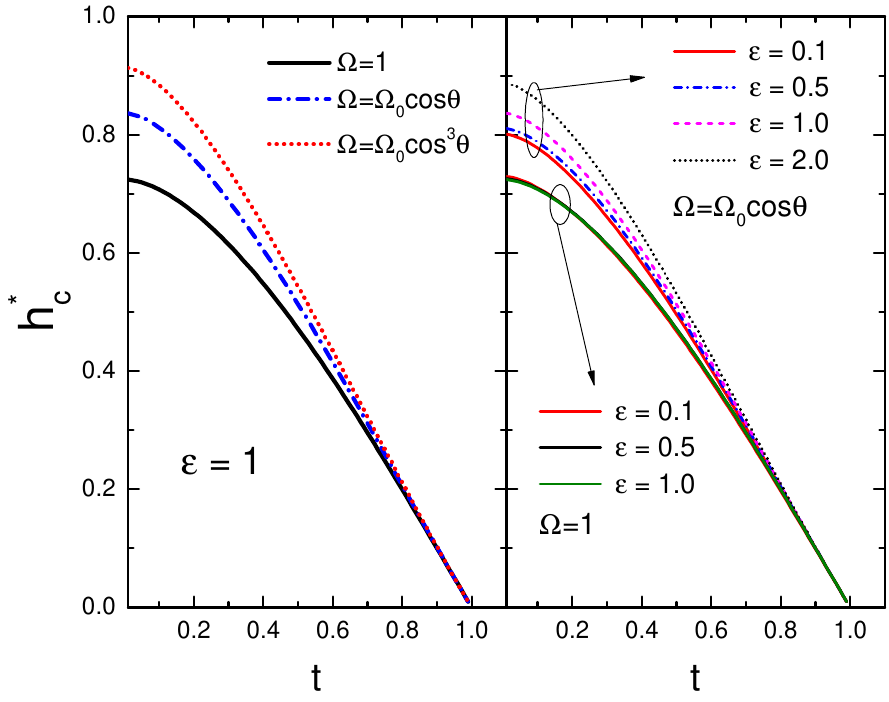}
\caption{(Color online) $h_c^*(t)$ for various Fermi surfaces   and   order parameters. Left panel: Fermi sphere and $\Omega=\sqrt{2n+1} \cos^n\theta$ with $n=0,1,3$. One sees that  equatorial nodes cause a substantial increase of  $h_c^*(0)$ and widen the domain of nearly linear behavior of $h_c^*(t)$. Note also a slight positive curvature for $n=1,3$.
Right panel: the lower group of three nearly coincident curves are for s-wave order parameter on a Fermi sphere and two prolate spheroids. The upper group is for $\Omega \propto\cos\theta$ showing clearly that  $h_c^*(0)$ increases with increasing $\epsilon$, the ratio of effective masses $m_c/m_{ab}$ (of squared spheroids semi-axes).
}
\label{f1}
\end{center}
\end{figure}

To study how the order parameter anisotropy affects $H_{c2,c}(T)$ and $h_c^*(0)$, we first consider the case of the Fermi sphere. We are interested   in $k_z$ dependent  order parameters, that on the Fermi sphere implies that $\Omega$ depends on the polar angle $\theta$.  
 We model equatorial nodes by setting  $ \Omega = \Omega_0 \cos^n  \theta $. 
Near the  ``equator" at $\theta=\pi/2$,  $|\Delta|$ behaves as $|\theta-\pi/2|^n$. Clearly, the bigger the power $n$, the wider  is the  equatorial belt where the order parameter is close to zero (we will call the power $n$ the ``node order").  It is readily shown that    
  \begin{eqnarray}
&&\Omega_0=\sqrt{2 n+1}\,,\qquad \left\langle\Omega^2\mu_c\right\rangle =   \frac{ 2(2n+1)}{ 4n^2+8n+3} \,,\nonumber\\
&& \left\langle\Omega^2\ln\mu_c\right\rangle =-{\bf C}-\psi\left(n+   3/ 2  \right) \,,
\label{Om^2ln mu-cos}
\end{eqnarray}
where $\psi$ is the digamma function.
  Hence, we have:
 \begin{eqnarray}
   h_c^*(0) =  \frac{7\zeta(3)}{4 } \frac{2n+1}{4n^2+8n+3} \,e^{ \psi\left(n+  3/2 \right) }.
\label{HW2}
\end{eqnarray}
     \begin{figure}[b]
\begin{center}
 \includegraphics[width=8.5cm] {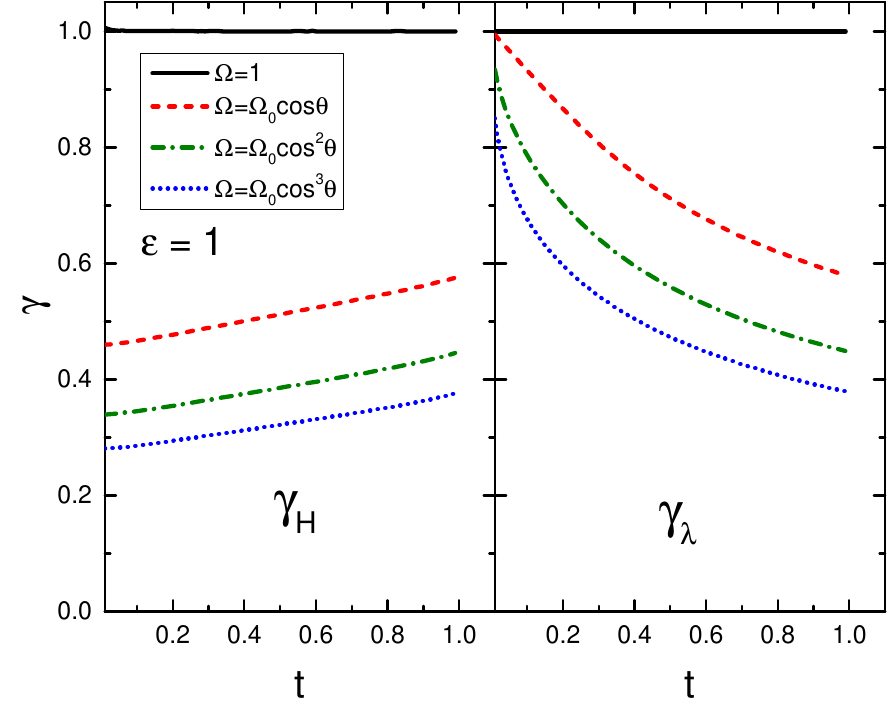}
\caption{(Color online) $\gamma_H=H_{c2,ab}/H_{c2,c}$ and $\gamma_\lambda=\lambda_c/\lambda_{ab}$ for the Fermi sphere and order parameters shown in the legend. Thus, the values of both $\gamma_H $ and $\gamma_\lambda $ are suppressed by equatorial line nodes; 
the suppression is stronger for higher  node orders. Besides, the nodes 
 cause the anisotropy of the upper critical field, $\gamma_H $, to increase on warming, whereas   $\gamma_\lambda$ decreases with increasing $T$, the feature reported, e.g., for Nd-1111 \cite{gam-lam(t)}. 
}
\label{f2}
\end{center}
\end{figure}
Hence, $h_c^*(0)$ increases with increasing $n$.  On the other hand, a larger $h_c^*(0)$ translates to a broader temperature range where $h(t)$ is close to being linear.  We then expect the curve $H_{c2,c}(T)$ to have an extended linear domain for increasing $n$.
  To check this statement we turn to   the full temperature dependence $h_c(t)$ which is found by solving numerically Eq.\,(\ref{eq-hc}). 
 The results are shown in the left panel of Fig.\,\ref{f1}. We estimate numerically  that $h_c^*(t)$ deviates from the straight line $h_c^{*\prime}(1)(t-1)$ by less than 1\% in the domain  $t>0.6 $ for $n=0$ (the s-wave), $t>0.4 $ for $n=1$, and $t>0.2 $ for $n=3$. Hence, increasing the node order causes ``straightening" of $H_{c2,c}(T)$  observed in pnictides \cite{Ni} and some other materials \cite{Lia}.
 
 Performing calculations for Fermi spheroids, one should evaluate properly  Fermi surface averages. Details of this procedure were worked out in \cite{MKM,KP-ROPP}.
Examples of $h_c^*(t)$ so obtained for a few values $\epsilon$, the squared ratio of the semi-axes, are given in the right panel of Fig.\,\ref{f1}. 

Similar to   Eq.\,(\ref{eq-hc}) for $H_{c2,c}(T)$, one can obtain an equation for $H_{c2,ab}(T)$, or directly for the anisotropy parameter $\gamma_H=H_{c2,ab}/H_{c2,c}$ \cite{KP-ROPP}. In fact, $\gamma_H(t)$ satisfies  Eq.\,(\ref{eq-hc}) in which, however, $h_c(t)$ is now known and $\mu_c$ should be replaced with 
     \begin{figure}[b]
\begin{center}
 \includegraphics[width=8.5cm] {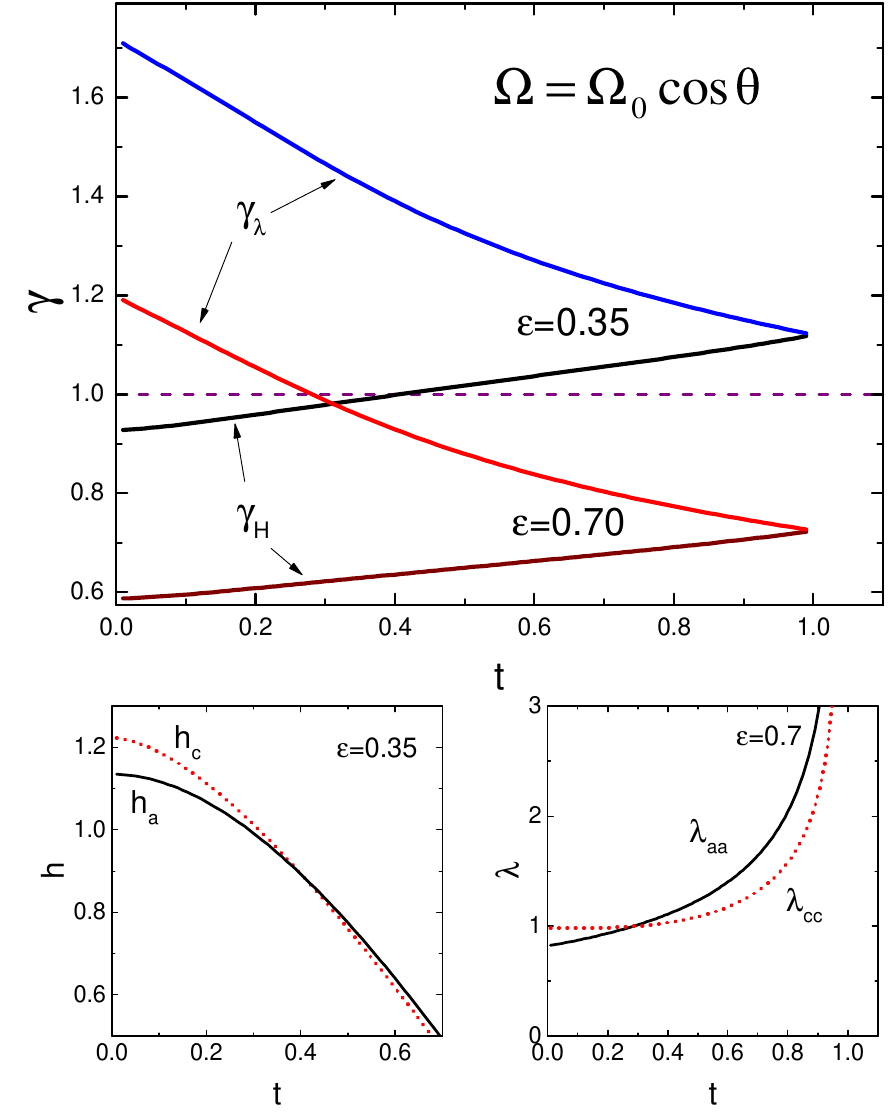}
\caption{(Color online) The upper panel: anisotropy parameters $\gamma_\lambda$ and $\gamma_H$ for the order parameter $\propto\cos\theta$ for two spheroidal Fermi surfaces  $\epsilon=0.35$ and 0.70. The lower left panel shows the crossing of $h_c(t)$ and $h_a(t)$ near $t^*=0.4$ for $\epsilon=0.35$. The lower right panel shows the crossing of $\lambda_{aa}(t)$ and $\lambda_{cc}(t)$ near $t^*=0.3$ for $\epsilon=0.70$; both $\lambda_{aa}(t)$ and $\lambda_{cc}(t)$ are normalized on $c/ev_0\sqrt{2\pi N(0)}$. Note the linear temperature  dependence of $\lambda_{aa}$ at low $T$s and a flat ``s-wave-like" behavior of $\lambda_{cc}$
}
\label{f3}
\end{center}
\end{figure}
\begin{eqnarray}
   \mu_b  =\frac{ v_x^2+\gamma_H^2v_z^2}{v_0^2} \,.
\label{muc}
\end{eqnarray}
The left panel of Fig.\,\ref{f2} shows $\gamma_H(t)$ for equatorial line nodes with $n=1,2,3$. One sees that for this type of nodes  on a sphere (i) $\gamma_H <1$,  i.e., $H_{c2,c} >H_{c2,ab}$ and  (ii) $\gamma_H $ {\it increases on warming}, the feature ubiquitous for the Fe-based materials.   
 
On the other hand, in most  materials of interest such as pnictides, the  Fermi surfaces are warped cylinders and $ H_{c2,ab}>H_{c2,c}$; $\gamma_H(t)>1$ but it is not large.   Qualitatively, one can model these Fermi surfaces as prolate spheroids, for which  it was shown that $\gamma_H >1$ for s-wave order parameters \cite{MKM,KP-ROPP}. Thus,  effect  of equatorial nodes on $\gamma_H$ is  the opposite to that of prolate Fermi surfaces.  It is of interest therefore to study  order parameters $\propto \cos^n\theta$ on prolate spheroids.
Figure \ref{f3} shows  examples   for  prolate spheroids with  $ \epsilon=0.35$ and 0.70 and the order parameter   $ \propto\cos \theta$. Remarkably,    $\gamma_H-1$ changes sign near $t^*\approx 0.4$ so that $h_a<h_c $ for $t<0.4$ and otherwise at higher temperatures. \\
     \begin{figure}[b]
\begin{center}
 \includegraphics[width=8.cm] {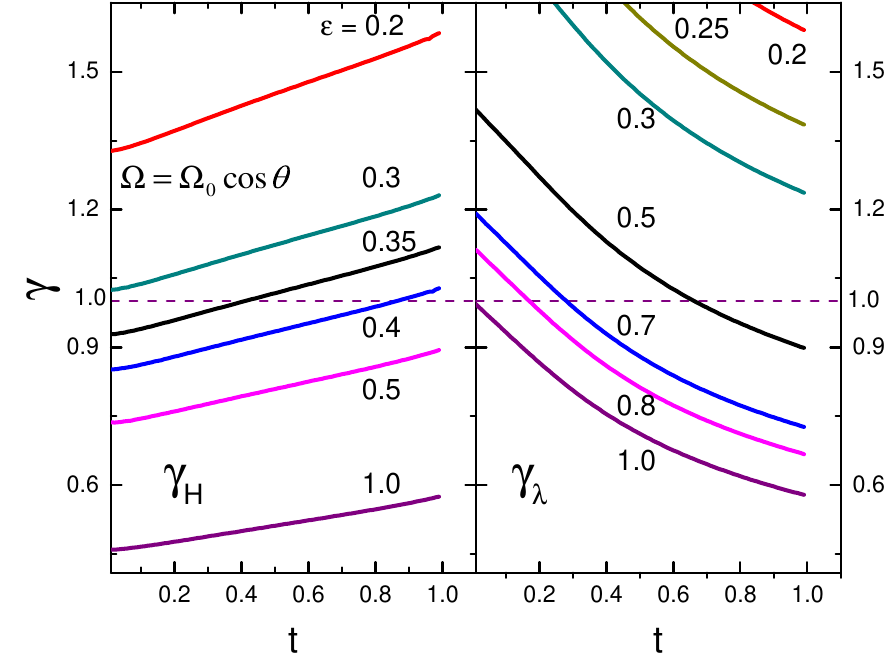}
\caption{(Color online) $\gamma_H(t) $ and $\gamma_\lambda(t)$ for the order parameter $\propto\cos\theta$ and the Fermi surface shapes $\epsilon$ indicated. One sees that $\gamma_H -1$ changes sign in the domain $0.3<\epsilon<0.45 $ whereas $\gamma_\lambda -1$ in the region $0.45<\epsilon<1$.
}
\label{f4}
\end{center}
\end{figure}
We now turn to the London penetration depth. The  inverse tensor of squared penetration depth for the general anisotropic clean case reads \cite{K2002,PK-ROPP}:
\begin{equation}
(\lambda^2)_{ik}^{-1}= \frac{16\pi^2 e^2 N(0) T}{c^2}\, \sum_{\omega} \Big\langle\frac{
\Delta^2v_iv_k}{\beta ^{3}}\Big\rangle \,.  
\label{lambda-tensor}
\end{equation}
Here $\Delta=\Psi \Omega $ and $\Psi(T)$ satisfies the self-consistency equation:
 \begin{eqnarray}
     - \ln t =\sum_{n=0}^\infty
\left (\frac{1}{n+1/2}-\left\langle\frac{ \Omega^2}{\sqrt{\psi^2\Omega^2+(n+1/2)^2}}\right\rangle\right) \,\qquad
\label{psi}
\end{eqnarray}
where $\psi=\Psi/2\pi T$. 

The  density  of states $N(0)$, Fermi velocities $\bm v$, and the order parameter anisotropy    $\Omega$ are the input parameters for 
evaluation of $\lambda_{aa}$ and $\lambda_{cc}$.
$N(0)$ is not needed if one is interested only in the anisotropy $\gamma_\lambda=\lambda_{cc}/\lambda_{aa}$:
\begin{eqnarray}
\gamma^2_\lambda&=& \frac{\lambda_{aa}^{-2}}{\lambda_{cc}^{-2}} 
=\frac{ \sum_n \left\langle
 \Omega ^2v_a^2/\eta^{3/2} \right\rangle}{  \sum_n \left\langle
 \Omega ^2v_c^2/\eta^{3/2} \right\rangle}\,, \nonumber\\
\eta&=&\psi^2\Omega^2+(n+1/2)^2\,.
  \label{gam-lam}
\end{eqnarray}
   It is easy to show that   Eq.\,(\ref{gam-lam}) gives
 \begin{eqnarray}
\gamma^2_\lambda(0) =\frac{  \langle  v_a^2\rangle} { \langle  v_c^2\rangle} \,,   \qquad\gamma^2_\lambda(T_c) =\frac{  \langle\Omega^2 v_a^2\rangle}
 { \langle\Omega^2 v_c^2\rangle}\,,
  \label{0,Tc}
\end{eqnarray}
as expected \cite{Gorkov,K2002}. We note that  $\gamma_\lambda=\gamma_H$ at $T_c$ because here the state is described by the anisotropic Ginzburg-Landau theory which contains only one ``mass" tensor responsible for both anisotropies.

The right panel of Fig.\,\ref{f2} shows $\gamma_\lambda $ evaluated with the help of 
Eq.\,(\ref{gam-lam}) for a Fermi sphere and $\Omega\propto\cos^n\theta$ with $n=0,1,2,3$. Hence, the equatorial line nodes cause $\gamma_\lambda(t) $ to {\it decrease on warming}, a behavior opposite to the increasing $\gamma_H(t) $ shown in the left panel. One also sees that the  two anisotropy parameters meet at $T_c$, thus confirming consistency of the analytic and numerical procedures for evaluation of two physically different quantities: the high field $H_{c2}(T)$ at the second order phase transition and the low field penetration depth $ \lambda (T)$. 
 
The combined effect of of the Fermi surface shape and of the order parameter $\Omega=\Omega_0\cos\theta$ on both $\gamma_\lambda $ and $\gamma_ H $ is shown on the upper panel of Fig.\,\ref{f3}. The Fermi surface  parameters $\epsilon$  are chosen to demonstrate interesting situations:  while $\gamma_\lambda>1 $ at all temperatures for $\epsilon=0.35$, the anisotropy $\gamma_ H $, being less than unity under $t^*\approx 0.4$, exceeds  1 above this temperature. Such a behavior has been recorded for Fe$_{1.14(1)}$Te$_{0.91(2)}$S$_{0.09(2)}$ \cite{FeTeS1} and for FeTeS \cite{FeTeS}. 

For $\epsilon=0.7$   we   have $\gamma_ H < 1$  at all temperatures, whereas
$\gamma_\lambda >1$ at $t^*<0.3$, but becomes less than unity above this temperature.  The transverse magnetization  of a material in the mixed state with such  $\gamma_\lambda $ placed in   field tilted relative to principal crystal directions should change sign at $t^*$ \cite{Tuominen}. The same is true for  the torque experienced by the crystal. In other words, the sign change of $\gamma_\lambda -1$ can be detected by measuring the sign and  angular dependence   of the transverse magnetization or torque \cite{K88, Farrell}.

  Fig.\,\ref{f4} shows that  temperatures $t^*$, at which $\gamma_H-1$ and $\gamma_\lambda-1 $ change sign, vary as functions of the Fermi surface shape  $\epsilon$: with increasing $\epsilon$ these temperatures grow if one goes to a ``less cylindrical" Fermi shapes. 
  
        \begin{figure}[h]
\begin{center}
 \includegraphics[width=7.5cm] {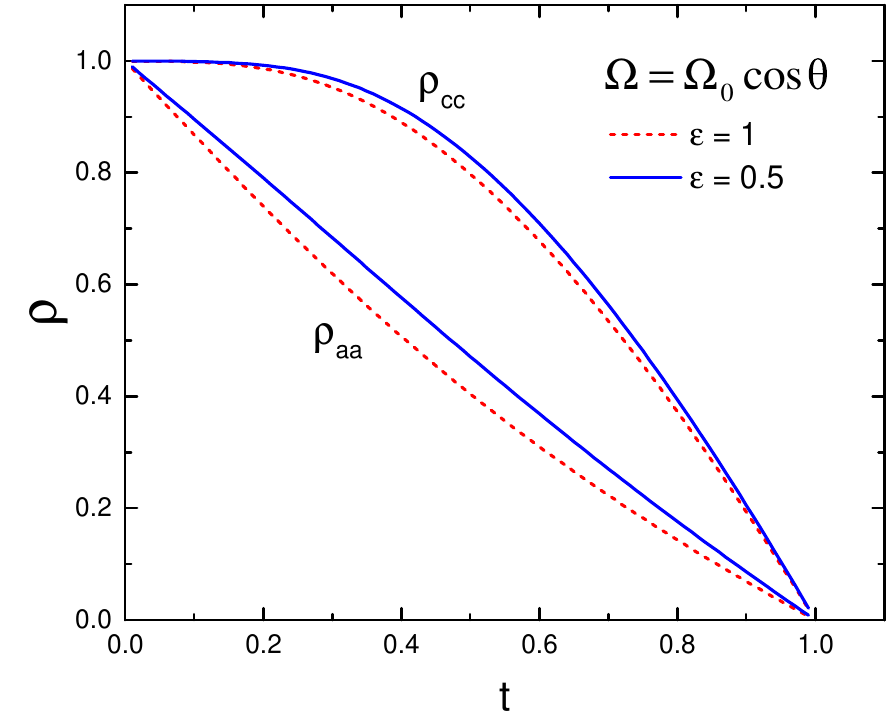}
\caption{(Color online) Superfluid densities  $\rho_{aa}=\lambda^2_{ aa}(0)/\lambda^2_{ aa}(t)$ (the lower curve) and $\rho_{cc}=\lambda^2_{cc}(0)/\lambda^2_{cc}(t)$.
}
\label{f5}
\end{center}
\end{figure}

A popular quantity in analysis of penetration depth data  is the   superfluid density defined as $\lambda^{-2}$ normalized on its  value at  $T=0$. This quantity for two principal directions  is plotted in Fig.\,\ref{f5} for an equatorial node, $\Omega=\Omega_0\cos\theta$, on a sphere and spheroid with $\epsilon=0.5$. Interestingly, the node presence results in $\rho_{aa}(t)$ qualitatively similar to the known d-wave linear low temperature behavior, whereas  direct numerical check shows that $\rho_{cc}-1\propto t^3$. In fact, this behavior has been discussed in \cite{Peter} considering properties of UBe$_{13}$.  

Concluding, we reiterate that despite profound simplifications, such as single-band ellipsoidal Fermi surface and  the order parameter with equatorial nodes, our model reproduces qualitative features of  anisotropic $H_{c2}(T)$ and $\lambda(T)$  often seen in real materials, notably  Fe-based superconductors. Fine details of Fermi surfaces  and order parameters  enter the theory of $H_{c2}(T)$ and $\lambda(T)$ only as averages over the Fermi  surface and thus do not justify formal complications  of taking them into account. Also, as far as $H_{c2}(T)$ and $\lambda(T)$ are concerned, the single- vs. multi-band scenarios give similar results as shown in our previous study \cite{KP-ROPP}.
Here we reproduced a number of features ubiquitous for Fe-based superconductors, origin of which up to now was not even questioned. In particular, we find that the equatorial line node causes an extended domain of nearly {\it linear} $H_{c2}(T)$, anisotropy of which {\it increases on warming}. By studying competing effects of equatorial nodes and of the Fermi surface anisotropy we find that, nearly cylindrical Fermi shapes notwithstanding, materials with equatorial nodes can be only weakly anisotropic. For certain combinations of material parameters both $\gamma_H-1$ and $\gamma_\lambda -1$ may {\it change sign on warming} so that $H_{c2,ab}<H_{c2,c}$ at low $T$s while  $H_{c2,ab}>H_{c2,c}$ at high $T$s. Similar situation may occur  for the anisotropy of the London penetration depth, which can be probed by torque or transverse magnetization measurements in large fields. We also find that the nodes in question cause different $T$ dependences of different components of the superfluid density tensor. These predictions call for   experimental verification.\\

 The authors are grateful to  M. Tanatar, A. Kaminsky, S. Bud'ko, V. Taufor, and P. Canfield for interest and discussions. This work was supported by the U.S. Department of Energy, Office of Science, Basic Energy Sciences, Materials Science and Engineering Division. The work was done at the Ames Laboratory, which is operated for the U.S. DOE by Iowa State University under contract  DE-AC02-07CH11358.

            \end{document}